Market laws

Çağlar Tuncay (*)

More than one billion data sampled with different frequencies from several financial instruments were investigated with the aim of testing whether they involve power law. As a result, a known power law with the power exponent around -4 was detected in the empirical distributions of the relative returns. Moreover, a number of new power law behaviors with various power exponents were explored in the same data. Further on, a model based on finite sums over numerous Maxwell-Boltzmann type distribution functions with random (pseudorandom) multipliers in the exponent were proposed to deal with the empirical distributions involving power laws. The results indicate that the proposed model may be universal.

It is known that power laws with various exponents can be interpreted as strong correlations in random data. Hence, that issue increasingly attracted attention from academic community as the amount and diversity of the data displaying power law behaviors increased with time. For example, the 1/f noise in electromagnetic signals from stars is one of the widely studied power law behaviors in science. Another example of power law with the power exponent around -1 can be found in human activities. The living world cities exhibit the mentioned power law behavior when they are organized in decreasing rank order with regard to the population sizes. That is known as Pareto distributions or Zipf's law [1].

Actually, the self-organized criticality approach was one of the early attempts to show that dynamical systems with spatial degrees of freedom naturally evolve into a self-organized critical point [2].

In the next important approach, referred to as highly optimized tolerance in literature, a mechanism for generating power law distributions was introduced, where the focus was on the systems which are optimized, either through natural selection or intentional design [3]. It was suggested that power laws in those systems are due to tradeoffs between yield, cost of resources, and tolerance to risks.

The reader is referred to the cited Newman papers in [1] and [4] for comprehensive treatments of power laws, and several points of comparison between self-organized criticality and highly optimized tolerance approaches.

Recently, Stanley's group investigated a large amount of data from several stock exchanges and they came up with an assertion saying that markets may be involving several power laws [5]. A particular form of those empirical laws is about the relative returns which will be outlined under the next heading.

In another important approach, Yakovenko's group studied the financial subjects after assuming that money is conserved in a closed economic system. Hence, by analogy with energy, the equilibrium probability distribution of money would follow Boltzmann-Gibbs law characterized by an effective temperature equal to the average amount of money per economic agent [6].

Moreover, Yakovenko group derived an analytical solution for the empirical probability distributions of relative returns as a function of time lags in Heston model of a geometrical Brownian motion with stochastic variance. They applied their formula to three major stock-market indices, NASDAQ, S&P500 and Dow-Jones from 1982 to 1999 inclusive, and concluded that the formula fitted a group of empirical probability distributions for various time lags between one day and one year inclusive.

Afterwards, they studied the probability distribution functions; precisely, cumulative distribution functions, of stock returns at mesoscopic time lags ranging from an hour to a month. They used the data from four shares of the NYSE; Microsoft Corporation (MSFT), Merck & Co. Inc. (MRK), International Business Machines Corporation (IBM) and Intel Corporation (INTC), and they found that the probability distribution of financial returns interpolates between exponential and Gaussian laws.

In summary, both Stanley and Yakovenko groups found that distributions of several financial parameters, such as relative returns follow power law. Moreover, exponential and Gauss distributions can be found in financial data. Furthermore, a function can be used for approximating several types of empirical distributions, but with different parameters. It means that, one may obtain the approximation of a specific type of empirical distributions from another approximation but after using a suitable set of parameters. That process is known as data collapsing. Hence, various distributions of derivatives can be collapsed onto other ones in terms of a formula. Those findings will be tested in further sections of this work.

*FINANCIAL DERIVATIVES*

Let us concern a time series for the historical values of a financial instrument r(τ;t) where τ is the sampling period of the data and the t is time. Thus, several relevant parameters can be defined as follows.

The plain returns (profits or losses) D(τ;t) are simply the differences of the closing prices of the successive time intervals of length τ

$$D(\tau;t) = r(\tau;t) - r(\tau;t-\tau) \quad . \qquad (1)$$

Hence, the relative returns (percentages of profits or losses) R(τ;t) can be written as

$$R(\tau;t) = D(\tau;t)/r(\tau;t-\tau)$$

$$= (\, r(\tau;t) - r(\tau;t-\tau)\,)/r(\tau;t-\tau) \quad . \qquad (2)$$

Similar expressions to those in Eqs. (1) and (2) can be written for the traded (tick-) volume. For example, if the number of executed buy or sell positions in a time interval of length τ is V(τ;t), then the variation in the traded volume W(τ;t) is

$$W(\tau;t) = V(\tau;t) - V(\tau;t-\tau) \quad . \qquad (3)$$

Thus, the relative change ω(τ;t) in the traded volume within the time term of length τ at t can be written as that

$$\omega(\tau;t) = W(\tau;t)/V(\tau;t-\tau) \quad . \qquad (4)$$

Another financial derivative object can be defined as the simultaneous ratios of the relative returns to the relative changes in the traded volumes per τ as that

$$S(\tau;t) = R(\tau;t)/\omega(\tau;t) \quad . \qquad (5)$$

In further sections, it will be tested whether the empirical distributions of D(τ;t), R(τ;t), W(τ;t), ω(τ;t) or S(τ;t) involve power law behaviors. It is worth noting that a power law with the power exponent around -4 was detected in the empirical distributions of R, from various stock exchanges in [5].

*DATA SET*

In order to achieve the above asserted aims a data set consisted of more than one billion raw data points being sampled with different periods from forex market is used [7]. The used sampling periods τ are 1, 5, 10, 60, 100, and 250 all in minute. The price steps are nearly equal to the rounded numbers of $(1/100)^{th}$ of a 1% of the prices (pip) which amounts to one or two digits in the fourth or fifth decimal places. Further details about the investigated data are given in Table 1.

*MODEL*

Let us concern a finite set of uniformly distributed random numbers $\{\lambda_i\}$ with $0 \le \lambda_i < 1$ and i=1,2, .., I. Let us define another set of uniformly distributed random numbers $\{\gamma_j\}$ with $0 \le \gamma_j < 1$ and j=1,2, .., J. which is obtained from the first set after organizing the elements in the increasing order; i.e.,

$$\lambda_i \rightarrow \gamma_j \text{ where one has } \gamma_j \le \gamma_{j+1} \qquad (6)$$

if J=I, and j=1, 2, …, J-1.

Then, let us define a new set of uniformly distributed and finite random numbers $\{B_j\}$ with $B_{min} \le B_j < B_{max}$ in terms of $\{\gamma_j\}$, and $B_{min}$ and $B_{max}$ as that

$$B_j = (B_{max} - B_{min})\, \gamma_j + B_{min} \quad . \qquad (7)$$

It is obvious that the random numbers $B_j$ are in increasing order due to Eq. (6).

Let us now assume that the sizes X in an agglomeration follow a function similar to Maxwell-Boltzmann type probability distribution function $O_{M\text{-}B}$ as that

$$O_{M\text{-}B}(B_j;X) = \exp(-XB_j) / Z \qquad (8)$$

where Z stands for the partition function

$$Z = \textstyle\sum_{j=1}^{J} \exp(-XB_j) \quad . \qquad (9)$$

In other words, Z is the total probability, and

$$\textstyle\sum_{j=1}^{J} O_{M\text{-}B}(B_j;X) = 1 \quad . \qquad (10)$$

*Uniformly distributed random variables*

Suppose that the integer J in Eqs. (6)-(10) is large enough to assume continuity for the random variable $B_j \rightarrow B$. Hence, the summation in Eq. (10) can be converted ( $\sum_{j=1}^{J} \rightarrow \int dB$ ) to the following integration

$$Z = \int (dB)\exp(-XB) \quad (11)$$

with the limits of $B_{max}$ and $B_{min}$. As a result,

$$Z = (\exp(-XB_{min}) - \exp(-XB_{max}))/X. \quad (12)$$

It is worth noting that Z depicts horizontal behavior for small sizes; i.e.,

$$Z \rightarrow (B_{max} - B_{min}) \text{ as } X \rightarrow 0 \quad (13)$$

due to l'Hôpital's Rule or since $e^{-X} \sim 1-X$ for small magnitudes (absolute values) of X. Moreover, Z in Eq. (12) can be approximated by a power function with the power exponent equals -1 for large sizes if further $B_{min}$ has a very small value and $B_{max} \gg B_{min}$. In this case, $\exp(-XB_{min}) \sim 1$ and $\exp(-XB_{max}) \sim 0$. Thus,

$$Z \sim 1/X . \quad (14)$$

Now, suppose that the $B_j$ in Eq. (7) are defined alternatively as

$$B_j = jB \quad . \quad (15)$$

In other words

$$B_{j+1} - B_j = ((j+1) - j) B = B$$

for $1 \leq j \leq J-1$ and

$$B_J - B_{J-1} = B \quad . \quad (16)$$

In the new situation Eq. (9) can be written as

$$\begin{aligned} Z' &= \sum_{j=1}^{J} \exp(-XB_j) \\ &= \sum_{j=1}^{J} \exp(-jXB) \\ &= \sum_{j=1}^{J} [\exp(-XB)]^j \\ &= \sum_{j=0}^{J} [\exp(-XB)]^j - 1 . \quad (17) \end{aligned}$$

The geometric series in Eq. (17) has the constant (common) ratio of $e^{-XB}$, and the result is

$$\begin{aligned} Z' &= [1/(1-\exp(-XB))] - 1 \\ &= (1-1+\exp(-XB))/(1-\exp(-XB)) \\ &= \exp(-XB)/(1-\exp(-XB)) \quad (18) \\ &= 1/(\exp(XB)-1) \quad . \quad (19) \end{aligned}$$

It is interesting that Z in Eq. (12) can be obtained from that Z' in Eq. (18) by treating that B in Eq. (16) as a uniform random variable.

Let us take the integral of Z' in Eq. (18)

$$Z'' = \int dZ' \quad (20)$$

which can be written as that

$$\begin{aligned} Z'' &= \int(dB)\exp(-XB)/(1-\exp(-XB)) \\ &= -(1/X)\int d(-XB)\exp(-XB)/(1-\exp(-XB)) \\ &= -(1/X) \int d\exp(-XB)/(1-\exp(-XB)) \\ &= (1/X) \int d(1-\exp(-XB))/(1-\exp(-XB)) \\ &= (1/X) \int d\ln(1-\exp(-XB)) \quad (21) \end{aligned}$$

with the integral limits of $(B_{max})$ and $(B_{min})$. As a result,

$$Z'' \sim [\exp(-XB_{min}) - \exp(-XB_{max})]/X \quad (22)$$

which is similar to that Z in Eq. (12); i.e., $Z'' \sim Z$.

The argument of the natural logarithm function in Eq. (21) is positive for finite values of X and B. Moreover,

$$\ln(1-e^{-XB}) \sim -e^{-XB}$$

$$\text{if} \quad 1-e^{-XB} \sim 1 \quad (23)$$

for both of the following cases, $B=B_{min}$ and $B=B_{max}$.

Another interesting result is that if the multiplication XB can be neglected in the sum with unity then the following approximation is meaningful

$$1/(\exp(XB)-1) \sim 1/(1-\exp(-XB)) \quad (24)$$

because $e^{XB} \sim 1 + XB$, and

$$1/(1+XB-1) = 1/(1-1-(-XB))$$

$$1/(XB) = 1/(XB) \quad . \quad (25)$$

It is worth underlining that Eq. (25) is valid only if $1 \gg XB$ which indicates power law behavior.

Another important note may be that X in Eq. (8) can be replaced by $\beta^n(Y-y)^n$. Moreover, the parameters y and β may satisfy the following equations

$$y = \int YZ(Y)dY \quad , \quad (26)$$

$$1/\beta = [\int(Y-y)^2Z(Y)dY]^{1/2} \quad . \quad (27)$$

Thus, they may be called the mean (expectation value of Y) and standard deviation, respectively. If, in this case, y=0 and β=1 then the corresponding distribution is called standardized. The reader may investigate the standardized distributions in a text book for probability theory, such as the one recommended in [8].

Generally speaking, one may use the following expression to approximate various empirical distributions

$$Z=[\exp(-B_{min}\beta^n(Y-y)^n) - \exp(-B_{max}\beta^n(Y-y)^n)]/[\beta^n(Y-y)^n] \quad (28)$$

and $Z'' \sim Z$. Moreover, $\beta^n$ can be absorbed in $B_{min}$ and $B_{max}$ in Eq. (28). Furthermore, one may write Eq. (19) as that

$$Z' = 1/[\exp(B\beta(Y-y) - 1] \quad . \qquad (29)$$

Note that Eq. (29) does not involve a power term in the exponential exponent, which is because of Eqs. (7), (15) and (16).

It is claimed that Eqs. (28) and (29) are useful for investigating various cumulative probability distributions of net wealth, those of tax returns, or Adjusted Gross Income of several countries which are discussed by the group of Yakovenko, but in terms of a different approach [6].

*Collapsing data*

The parameters $B_{min}$ and $B_{max}$ may well attain different values in different distributions. For example, they may depend on the sampling period $\tau$. With the aim of testing that possibility, the following assumptions will be made;

$$B_{min} = 0$$

and $$B_{max} = C/\tau^{\alpha}, \qquad (30)$$

where C and $\alpha$ are real constants, and $\beta$ is absorbed in C. The value for the $\alpha$ in Eq. (30) will be defined empirically in this work.

As a result, that Z in Eq. (28) can be written under the recent assumptions as

$$Z = (1 - \exp(-\beta^n Y^n C/\tau^{\alpha})) / Y^n. \qquad (31)$$

if y=0.

Now, if $\Psi(Y)$ represents the empirical distributions of a derivative Y or the magnitudes of them, then those distributions can be collapsed on the number unity in terms of the proposed model as follows

$$\Psi' = \Psi\beta^n Y^n + \exp(-\beta^n Y^n C/\tau^{\alpha}) \propto 1 \qquad (32)$$

where the symbol $\propto$ is used instead of the equality sign since Eq. (32) involves empirical distributions. The previous claim can be reformulated as the following. For a given empirical distribution $\Psi(D)$, $\Psi(R)$, $\Psi(W)$, $\Psi(\omega)$ or $\Psi(S)$, there exists a positive number (n) for which Eq. (32) holds.

Another collapsing method may be that

$$\Psi(Y)/Z(Y) \propto 1 \qquad (33)$$

and this finishes the model. Results of the applications will be presented in the following section. The last section is devoted to discussion and conclusion.

*APPLICATIONS AND RESULTS*

Primarily, each of the downloaded one-year-long time series (see Table 1) sampled with different frequencies was investigated for whether they are problematic due to a reason. In the meantime, they were plotted in the OHLC-V (open, high, low, and close, and volume) fashion and compared to the graphs of the same instruments supplied by at least two independent sources, such as www.investing.com and a forex house.

Afterwards, the derivatives D, R, W, $\omega$, and S were computed from each yearly segment with a time span of nearly 375,000 trading minutes. Then, the empirical distributions of those derivatives from all of the investigated instruments sampled with different $\tau$ were plotted with various bin sizes (not showed). Summations of the event frequencies of those yearly distributions with specific bin sizes were used to obtain the total event frequencies per instrument.

i- Bilateral symmetry and power law

The observations made on inspection of the plots mentioned in the previous paragraph can be outlined as the following. Distributions are unimodal, and the modes occur at around zero. They display almost bilateral symmetry about the vertical axis. This means that the event frequencies are nearly the same for the studied derivatives or negatives of those.

The magnitudes of the means, as defined in Eq. (26), are very close to zero as exemplified for the available data from several instruments in Table 2. Thus, the means can be taken practically zero since their magnitudes are smaller than the bin size used for the computed data.

Moreover, the yearly distributions of the positive values of the D, R, W, $\omega$ and S with $\tau$=1 minute usually displayed almost power laws with indices around -4, -4, -2, -2, and -2, respectively. Furthermore, the mentioned power law distributions showed nearly similar behaviors when the studied parameters are multiplied by -1. This means that, if a power law can be detected in the empirical distributions of a derivative and negative of that then the same power law can be detected in the distributions of the magnitudes. However, no clear power law behavior could be detected in the empirical distributions of the relative returns from several instruments. Another important note is that no clear power laws could be observed in the distributions of several studied parameters, such as W and $\omega$ from lightly traded instruments or from early years when the related instruments were less popular.

Bilateral symmetric distributions of the $R(\tau=1min;t)$ from the available data for two groups of forex instruments are exemplified in Figures 1 (a) and (b). On inspection of those figures it can be observed that the exemplified distributions are almost bilateral symmetric, unimodal and the modes occur at around zero. Those results can be attributed to the following reasons.

The profits or losses ($0<D(\tau;t)$ or $0>D(\tau;t)$, respectively) occur with nearly the same event frequencies in lengthy time intervals. Therefore, the empirical distributions for the relative returns $R(\tau;t)$ are expected to depict almost bilateral symmetry if the $D(\tau;t)$ distribute with almost bilateral symmetry and $r(\tau;t)$ varies smoothly with time. Moreover, if the distributions of $D(\tau;t)$ follow a power law with the power exponent close to minus four then also the distributions of $R(\tau;t)$ follow a power law with the power exponent close to minus four but only for horizontal trends in prices. It is because the $r(\tau;t)$ terms can be taken constant in the simultaneous ratios $D(\tau;t)/r(\tau;t)$ thus, $D(\tau;t) \propto R(\tau;t)$.

Distributions of the $R(\tau=1\text{min};t)$ from the available data of several forex instruments depicting the known power law with the index of -4 or those from several other instruments with no clear power law are exemplified in Figures 2 (a) and (b), respectively. On inspection of Figs. 2 (a) and (b) it can be observed that the recently mentioned power law cannot be treated as universal because it occurs not in all of the investigated time intervals.

It is worth underlining that the symmetry relations mentioned for $D(\tau;t)$ and $R(\tau;t)$ in the previous three paragraphs are detected in the distributions of $W(\tau;t)$ and $\omega(\tau;t)$, too. Here, the basic reason is that the $W(\tau;t)$ distribute symmetrically and $V(\tau;t)$ vary smoothly but only if they are obtained from heavily traded instruments. Further on, a number of extra power law behaviors are explored in the distributions of the simultaneous ratios ($R^{\theta}/\omega^{\phi}$) with various values for the integers $\theta$ and $\phi$. However, further dealing with symmetric distributions or the algebra for ratios of the power law distributions is not relevant to this paper. The interested reader may be referred to the papers cited in [9] for the differences or ratios of distributions.

ii- Distributions in rank order

It is known that if the power exponent of a power law distribution is less than or equal to minus two then that of the cumulative distribution of the same random variable is greater by one. Moreover, the cumulative distribution is simply proportional to the rank of that variable [1]. Thus, the power exponents for the rank/frequency plots can be expected to attain the values close to -3, -3, -1, -1, and -1 for the D, R, W, ω, and S, respectively. With the aim of testing that claim, the event frequencies of the D, R, -R, W, ω, and S from the EURUSD in 2012 with $\tau=1$ min are organized in the decreasing rank order and presented in Figure 3. The reader may utilize the Supplemental Material I, which is available for download from the link given in [10], to study various relevant rank/frequency plots.

On inspection of Fig. 3 (and the figures displayed in the Supplemental Material I) it can be observed that the power exponents of the rank distributions of the derivatives D, -R, and R with $\tau=1$min are commonly around -3. Note that those exponents come out around -3 also for the absolute values of the same derivatives as can be found in Supplemental Material I. Moreover, the power exponents for the same derivatives or magnitudes of those are commonly around -4 in the empirical distributions which were discussed under the previous heading. Furthermore, the computed data for R and -R follow each other closely which is due to the almost bilateral symmetry observed in the related empirical distributions. It is worth underlining that similar plots are obtained from the asserted derivatives or negative of them which were computed using the data sampled in different years. However, important alterations occurred in the distributions as a result of the varied sampling periods. For example, less clear power law behaviors occurred in the data obtained with long sampling periods as exemplified in Figure 4.

Therefore, the power law behaviors are sensitive to the sampling periods. Moreover, the power exponents in the empirical distributions are known to depend on the utilized bin sizes [11]. Further on, disregarding a number of contrastive data may cause unreliable estimations of the related power exponents [1].

Hence, the known processes for estimating power exponents of the power laws lack confidence. In other words, power law approaches for investigating a number of empirical distribution types, such as the distributions of financial derivatives are open to dispute, if not totally meaningless.

iii- Collapsing

The Supplemental Material II [12] is used to obtain several simulations exhibited under this heading, such as those in Figures 5 - 8. On inspection of those figures it can be observed that the proposed model is capable of simulating various distributions involving power law regimes with positive or negative power exponents. It is obvious that the fluctuations at the tails of the simulations obtained using Eq. (9) after assuming $X \rightarrow Y^n$ (with y=0) can be decreased by increasing the value of J in the same equation.

Now, let us test whether the novel model is useful for approximating the very financial distributions of the investigated derivatives with various $\tau$. It is known that least squares fitting is one of the most favored processes for approximating discrete distributions with continuous functions. But instead, the above described data collapsing processes will be followed in this section. Another aim is to test whether it is possible to obtain an approximate function for the empirical distributions of a specific financial parameter obtained from a given instrument with a certain sampling period in terms of another approximate function for the empirical distributions of the same or different financial parameter obtained from the same or different instrument with the same or different sampling period. It should be noted that the distributions of the $R(\tau=1min;t)$ calculated from the available data of EURUSD and XAGUSD are used to exemplify the application of Eq. (32) in the Supplemental Material III (a) and (b), respectively [13].

Let us consider $\Psi(R)$ from the available data specific to a financial instrument. The model parameters will be n=4 and $B_{min}=0$. The $B_{max}$ will be defined for each instrument as the value of the mode of the empirical distributions $\Psi$. However, the extreme values $B_{max}$ are found to depend on the sampling periods ($\tau$) as exhibited in Figures 9 (a) and (b), Figure 10, and Table 3. On inspection of Table 3 and Fig. 10 it can be observed that $B_{max}$ for the investigated forex instruments commonly follow a clear power law with the common power exponent around 1.4 as a function of the sampling frequencies ($1/\tau$) in Hz. Hence, Eq. (30) can be written as that $B_{max} \propto 8.5/\tau^{1.4}$ but only approximately. The actual relations used for the collapsing applications are;

$$(B_{max})_{EURUSD} \propto 8.99124/\tau^{1.4481} \quad (34)$$

and $$(B_{max})_{XAGUSD} \propto 7.97702/\tau^{1.4029} \quad (35)$$

where $\tau$ is in seconds.

The result of the data collapsing process following Eq. (32) and using the parameters in Table 2 is exemplified for the eleven-year-data from EURUSD and XAGUSD in Figures 11 (a) and (b), respectively. On inspection of Figs. 11 (a) and (b), and the applications presented in Supplemental Material III (a) and (b) it can be observed that the proposed model is reliable for precise approximations of the empirical distributions of the $R(\tau;t)$ from EURUSD and XAGUSD. It is worth noting that each of the distributions plotted in Figs. 11 (a) and (b) involve nearly 4.125 million computed data.

Also, one can test the reliability of Eq. (33). Results of the collapsing processes for the $R(\tau;t)$ from the available data of EURUSD and XAGUSD are displayed in Figures 12 (a) and (b), respectively. The reader is referred to the Supplemental Material III (c) and Material III (d) for investigating other examples of the current collapsing process [14].

On inspection of Figs. 11 (a) and (b), and Figs. 12 (a) and (b) it can be observed that the recent results support each other. Moreover, it is obvious that the applications of Eqs. (32) and (33) are not limited to distributions involving power law behaviors with power exponent around -4. They can be used for the distributions of $S(\tau,t)$ from various forex instruments, too. The reader may note that the rank distributions of the $S(\tau=1\ min,t)$ involve power law behavior with the power exponent around -1 as mentioned several times, previously. For that reason, Eq. (29) can be used to approximate the rank distributions of the $S(\tau=1\ min,t)$ as exemplified in Figures 13 (a) and (b). The reader may utilize the Supplemental Material IV for various new simulations for the most recently mentioned approximating process [15].

iv- Mixed behaviors

The above presented results indicate that the proposed model is capable of approximating various mixed behaviors in the empirical distributions. For example, the R- and S-distributions commonly involve horizontal and power law behaviors near the modes and tails, respectively. Moreover, they can be approximated in terms of the function defined in Eq. (28).

Another type of mixed distributions may involve Gaussian, exponential and power law behaviors at a time. For example, the cumulative probability distributions of net wealth in the UK and those of tax returns for USA in 1997, both of which involve different regimes are exhibited in Figures 14 (a) and (b), respectively.

A simulation in terms of Eq. (28), and with n=1, y=7, $B_{min}=0$, $B_{max}=1$, and $\beta=1$ can be generated to cover the power-law and exponential regimes in the same distribution as exemplified in Figures 15 (a) and (b).

The reader may use the Supplemental Material V for generating various simulations exhibiting different regimes at a time [16].

*DISCUSSION and CONCLUSION*

It is empirically predicted that frequency distributions of various financial parameters, such as simple and relative changes of prices and traded volumes, and their simultaneous ratios involve mixed behaviors. They depict horizontal behaviors around the modes, almost power law decays shaping the tails as well. Moreover, the whole body of those behaviors can be well approximated in terms of the proposed three-parameter model, one of which may be taken zero. Another parameter stands for the mode value, and the third one stands for the power exponent of the related power law behavior. Hence, the model can be used to obtain an approximating function for a specific type of empirical distributions in terms of that for another type. Therefore, empirical distributions of several financial parameters obtained from various instruments using various sampling periods can be collapsed onto one another or an arbitrary real number, as well.

*Similarities and dissimilarities between physics and finance*

The above presented theory and its results indicate that the proposed model works, especially if the uniform real random variables follow exponentially decreasing probabilities. For example, it is known that the time rate of nuclear disintegrations in a specific radioactive material is constant. Thus, the transmutations occur exponentially in time. Suppose further that a number of atom types are present in the material, and the disintegration rates attain uniformly distributed random values. Then, the radioactive abundances will follow Eq. (12) but only after replacing the X with time t. Moreover, the rates may vary following Eqs. (15) and (16); that is, with constant differencing. In that case, the distributions will follow that Z' in Eq. (19) which is similar to Bose-Einstein occupancies in quantum statistics. The recently mentioned differencing values occur due to the price steps which may well be different in different time intervals for a forex market. More specifically, the prices and thus the price steps are defined as an integer times a small fraction ($\sim 10^{-5}$) of unit of the related currency. In that case, the result for the distributions is that Z" in Eq. (22).

It should be noted that the binning sizes used for obtaining the empirical distributions must be around the price steps. If those sizes are smaller than the price steps then several bins will be empty. On the other hand, large bin sizes may cause passing over various relevant information about the distributions. Moreover, number of the executed data must be near the reciprocal of the related price step for obtaining full bins. Otherwise, the results may be misleading.

Conclusively, natural and human-made systems may depict similar behaviors if the underlying mechanisms for the related event frequency distributions occur similarly. That may explain why a diversity of empirical distribution types involve almost power law behaviors, and why those laws are ubique in physics, finance and several more disciplines.

With the aim of explaining the financial power laws it can be argued that each of the numerous traders follow a specific strategy or an agent follows a number of different strategies equally well, while trading a voluminous instrument. For example, the stop-loss or take-profit levels (in percentages) may be diverse. Moreover, those levels may be uniformly distributed around the prices at a time. Furthermore, the number of those levels may be increasing exponentially as they come close to the current prices. Hence, the realizations of the fails and successes may be distributed in a bilateral symmetric manner in event frequencies per unit of capital. Due to those reasons, the model parameters y and n in Eq. (28) are expected to be equal to zero and an even integer, respectively.

Specifically speaking, the buying and selling processes occur with almost equal probabilities within a considerable time interval normally; i.e., away from transition eras. Also, the probabilities for profits and losses are almost equal in the same interval. Therefore, the growth and regression of the used capitals are reversible processes with almost equal magnitudes for the process rates. Thus, the parameters defined in Eqs. (1)-(5) distribute in a bilateral symmetric manner involving almost power law behaviors with the power exponents in even integers. However, nuclear disintegrations are not reversible and they depict almost power law behavior with odd integer which is unity. Note that this situation can be observed in various phenomena, such as 1/f noise and population distributions of the living cities. In other words, the occurrence of Pareto distribution or Zipf law in urban may be attributed to an irreversible process, such as population growth.

*Conclusion*

Distributions of several financial parameters, such as relative returns involve power law behaviors. That conclusion is in agreement with the results presented in [5]. Moreover, power law, exponential and Gauss behaviors can be observed in different parts of an investigated distribution [6]. Furthermore, a function can be used for approximating several types of empirical distributions, but with different parameters. It means that one may obtain the approximation of a specific type of empirical distributions from another approximation but after choosing a suitable set of model parameters. Thus, various distributions of derivatives can be collapsed onto other ones in terms of a formula. Note that the formula proposed in Eqs. (28) and (29) is simple with regard to that used in [6]. This author challenges that there may be a class of equivalent formulas for collapsing the financial data.

## APPENDIX

Supplemental Material I is composed of two Microsoft Office Excel sheets. The historical data sampled with 1 minute period from the EURUSD in different years can be found in Sheet2 and the distributions of the derivatives D, R, W, ω and S in Eqs. (1)-(5) and those of their magnitudes in decreasing rank order can be found in Sheet1 which are indicated in yellow and blue, respectively. The data in the columns A and B of Sheet1 can be deleted and new data can be copied from the Sheet2 and pasted in a special manner for the values within the A and B columns in Sheet1. Depending on the configuration of Excel program used by the reader, the new computations take some time and the graphs get updated.

Supplemental Material II involves the simulations of the proposed model in Eq. (29) with y=0 in logarithmic and log-log scales in Sheet1 and Sheet2, respectively. The reader will enter a real number in A1 cell in a sheet to define the power exponent and wait for a while (as in the previous case) for the updates to be completed.

Supplemental Material III (a) and (b), and (c) and (d) can be used with the aim of investigating the results of the performed data collapsing process for EURUSD and XAGUSD in terms of Eqs. (32), and for EURUSD and XAGUSD in terms of (33), respectively.

Supplemental Material IV involves various applications of that Z' in Eq. (19). The data in the A and B columns can be copied down from the Sheet2 of the Supplemental Material I, as described above. The parameters D, R, W, ω and S will be computed automatically. The parameters in Y3 and Y4 cells can be varied for obtaining a diversity of results.

Supplemental Material V involves simulations for the mixed behaviors of exponential and power law. The value of the b in the cell of E4 can be used to multiply the values of the ($B_{min}$) and ($B_{max}$) instead of changing them manually.

**Acknowledgments**
This paper is dedicated to the etched memories of Dietrich Stauffer who was always eager for friendly discussions. The author is thankful to Victor Yakovenko for his permission for reproduction of two figures used in Figs. 14 (a) and (b). A special thank goes to Mr. Murat Duman from Orta Doğu Teknik Üniversitesi Bilgi İşlem Daire Başkanlığı, who assisted recording and sampling data.

REFERENCE LIST

(*) caglart@metu.edu.tr

1. Newman M., E., J., Power laws, Pareto distributions and Zipf's law, Contemporary Physics, **46**, 323-351 (2005). (arXiv:cond-mat/0412004). Also, Clauset A., Shalizi C. R., and Newman M., E., J., Power-Law Distributions in Empirical Data, SIAM Rev., **51**, 661–703 (2009). Also, Reed W., J., The Pareto, Zipf and other power laws, Economics Letters **74**, 15–19 (2001).

2. Bak, P., Tang, C. & Wiesenfeld, K., Self-Organized Criticality: An Explanation of 1/f Noise, **59**, 381-384 (1987).

3. Carlson, J., M., & Doyle, J., Highly optimized tolerance: A mechanism for power laws in designed systems, Phys. Rev., **60**, 1412-1427 (1999). (arXiv:cond-mat/9812127)

4. Newman, M., The power of design, Nature, **405**, 412-413 (2000).

5. Liu Y., et al. The statistical properties of the volatility of price fluctuations. Phys. Rev. E 60, 1390–1400 (1999). (arxiv:cond-mat/9903369). Also, Gabaix X et al., A theory of power-law distributions in financial market fluctuations. Nature, 423, 267–270 (2003).

6. Dragulescu A. A. and Yakovenko V. M., Statistical mechanics of money, The European Physical Journal **B 17**, 723-729 (2000). Also, Silva A. C. and Yakovenko V. M., Comparison between the probability distribution of returns in the Heston model and empirical data for stock indexes, Physica **A 324**, 303-310 (2003). Also, Silva A. C., Prange R. E. and Yakovenko V. M, Exponential distribution of financial returns at mesoscopic time lags: a new stylized fact, Physica **A 344**, 227-235 (2004). Also, Yakovenko V., M., Econophysics, Statistical Mechanics Approach to, arXiv:0709.3662, (2008).

7. URL: http://www.tickstory.com

8. See pages 69-75 in Haigh J., Probability Models (Second Edition, Springer-Verlag London 2002, 2013).

9. Fama E., F., and Roll R., Some Properties of Symmetric Stable Distributions, Journal of the American Statistical Association **63**, 817-836, (1968). Ramberg J., S., and Bruce W., S., An Approximate Method for Generating Symmetric Random Variables, Communications of the ACM (Association for Computing Machinery), **15**, 987-990 (1972). Dynkin E., B., Mandelbaum A., Symmetric Statistics, Poisson Point Processes, and Multiple Wiener Integrals, The Annals of Statistics, **11**, 739-745, (1983). Gregory L., P., et. al., Computational Methods for measuring the Difference of Empirical Distributions, Amer. J. Agr. Econ. **87**, 353–365 (2005).

10. Supplemental Material I; worksheet, 196,205kB, URL: https://drive.google.com/file/d/1HfMVfyuUl89mSiBkmGrp2GHLb6ga1XA8/view?usp=sharing

11. Virkar Y., and Clauset A., Power-Law Distributions in Binned Data, The Annals of Applied Statistics, **8**, 89–119 (2014).

12. Supplemental Material II; worksheet, 1,570kB, URL: https://drive.google.com/file/d/1sF1eudcsGbrpxBUBMaUern4_acjHkETb/view?usp=sharing

13. Supplemental Material III (a) and (b); worksheets, 12,111kB and 12,934kB, respectively, URLs: https://drive.google.com/file/d/1OfVpZvjramHGozFmSVI72_sSbH6-VjTd/view?usp=sharing and https://drive.google.com/file/d/1OaBg65n8PSAU2mkrwLmIKVSzW_t5OsBG/view?usp=sharing, respectively.

14. Supplemental Material III (c) and (d) ; worksheets, 10,581kB and 11,009kB, respectively, URLs: https://drive.google.com/file/d/1kXGxnQHOxdBrdaEl6jpQfnq0WyEzbcBl/view?usp=sharing and https://drive.google.com/file/d/1kf1OmvnV8bp0j2kGCRJ4exdC4Nfv5rJI/view?usp=sharing, respectively.

15. Supplemental Material IV; worksheet, 205,187kB, URL: https://drive.google.com/file/d/1_1lCWg1Cz7tijMJNgPddiE8OMs5MPWyn/view?usp=sharing.

16. Supplemental Material V; worksheet, 14,610kB, URL: https://drive.google.com/file/d/1tD_n2ItQ_rwjCw5HHqvLA3tuI0PFh0Pn/view?usp=sharing.

TABLES

| Symbol | Name | year (20 . .) 3 | 4 | 5 | 6 | 7 | 8 | 9 | 10 | 11 | 12 | 13 | 14 |
|---|---|---|---|---|---|---|---|---|---|---|---|---|---|
| AUDCAD | Australian Dollar/Canadian Dollar | - | - | - | + | + | + | + | + | + | + | + | + |
| AUDCHF | Australian Dollar/Swiss Franc | - | - | - | + | + | + | + | + | + | + | + | + |
| AUDJPY | Australian Dollar/Japanese Yen | + | + | + | + | + | + | + | + | + | + | + | + |
| AUDNZD | Australian Dollar/New Zealand Dollar | - | - | - | + | + | + | + | + | + | + | + | + |
| AUDSGD | Australian Dollar/Singapore Dollar | - | - | - | - | + | + | + | + | + | + | + | + |
| AUDUSD | Australian Dollar/US Dollar | + | + | + | + | + | + | + | + | + | + | + | + |
| CADCHF | Canadian Dollar/Swiss Franc | - | - | - | + | + | + | + | + | + | + | + | + |
| CADHKD | Canadian Dollar/Hong Kong Dollar | - | - | - | - | + | + | + | + | + | + | + | + |
| CADJPY | Canadian Dollar/Japanese Yen | - | + | + | + | + | + | + | + | + | + | + | + |
| CHFJPY | Swiss Franc/Japanese Yen | + | + | + | + | + | + | + | + | + | + | + | + |
| CHFPLN | Swiss Franc/Polish Zloty | - | - | - | - | + | + | + | + | + | + | + | + |
| CHFSGD | Swiss Franc/Singapore Dollar | - | - | - | - | + | + | + | + | + | + | + | + |
| EURAUD | Euro/Australian Dollar | - | - | + | + | + | + | + | + | + | + | + | + |
| EURCAD | Euro/Canadian Dollar | - | + | + | + | + | + | + | + | + | + | + | + |
| EURCHF | Euro/Swiss Franc | + | + | + | + | + | + | + | + | + | + | + | + |
| EURDKK | Euro/Danish Krone | - | + | + | + | + | + | + | + | + | + | + | + |
| EURGBP | Euro/Pound Sterling | + | + | + | + | + | + | + | + | + | + | + | + |
| EURHKD | Euro/Hong Kong Dollar | - | - | - | - | + | + | + | + | + | + | + | + |
| EURHUF | Euro/Hungarian Forint | - | - | - | - | + | + | + | + | + | + | + | + |
| EURJPY | Euro/Japanese Yen | + | + | + | + | + | + | + | + | + | + | + | + |
| EURMXN | Euro/Mexican Nuevo Peso | - | - | - | - | + | + | + | + | + | + | + | + |
| EURNOK | Euro/Norwegian Krone | - | + | + | + | + | + | + | + | + | + | + | + |
| EURNZD | Euro/New Zealand Dollar | - | - | - | + | + | + | + | + | + | + | + | + |
| EURPLN | Euro/Polish Zloty | - | - | - | - | + | + | + | + | + | + | + | + |
| EURRUB | Euro/Russian Ruble | - | - | - | - | + | + | + | + | + | + | + | + |
| EURSEK | Euro/Swedish Krona | - | + | + | + | + | + | + | + | + | + | + | + |
| EURSGD | Euro/Singapore Dollar | - | - | - | - | + | + | + | + | + | + | + | + |
| EURTRY | Euro/Turkish Lira | - | - | - | - | + | + | + | + | + | + | + | + |
| EURUSD | Euro/US Dollar | + | + | + | + | + | + | + | + | + | + | + | + |
| EURZAR | Euro/South Africa Rand | - | - | - | - | + | + | + | + | + | + | + | + |
| GBPAUD | Pound Sterling/Australian Dollar | - | - | - | + | + | + | + | + | + | + | + | + |
| GBPCAD | Pound Sterling/Canadian Dollar | - | - | - | + | + | + | + | + | + | + | + | + |
| GBPCHF | Pound Sterling/Swiss Franc | + | + | + | + | + | + | + | + | + | + | + | + |
| GBPJPY | Pound Sterling/Japanese Yen | + | + | + | + | + | + | + | + | + | + | + | + |
| GBPNZD | Pound Sterling/New Zealand Dollar | - | - | - | + | + | + | + | + | + | + | + | + |
| GBPUSD | Pound Sterling/US Dollar | + | + | + | + | + | + | + | + | + | + | + | + |
| HKDJPY | Hong Kong Dollar/Japanese Yen | - | - | - | - | + | + | + | + | + | + | + | + |
| MXNJPY | Mexican Nuevo Peso/Japanese Yen | - | - | - | - | + | + | + | + | + | + | + | + |
| NZDCAD | New Zealand Dollar/Canadian Dollar | - | - | - | + | + | + | + | + | + | + | + | + |
| NZDCHF | New Zealand Dollar/Swiss Franc | - | - | - | + | + | + | + | + | + | + | + | + |
| NZDJPY | New Zealand Dollar/Japanese Yen | - | - | - | + | + | + | + | + | + | + | + | + |
| NZDSGD | New Zealand Dollar/Singapore Dollar | - | - | - | - | + | + | + | + | + | + | + | + |
| NZDUSD | New Zealand Dollar/US Dollar | + | + | + | + | + | + | + | + | + | + | + | + |
| SGDJPY | Singapore Dollar/Japanese Yen | - | - | - | - | + | + | + | + | + | + | + | + |
| USDBRL | US Dollar/Brazilian Real | - | - | - | - | + | + | + | + | + | + | + | + |
| USDCAD | US Dollar/Canadian Dollar | + | + | + | + | + | + | + | + | + | + | + | + |
| USDCHF | US Dollar/Swiss Franc | + | + | + | + | + | + | + | + | + | + | + | + |
| USDDKK | US Dollar/Danish Krone | + | + | + | + | + | + | + | + | + | + | + | + |
| USDHKD | US Dollar/Hong Kong Dollar | - | - | - | - | + | + | + | + | + | + | + | + |
| USDHUF | US Dollar/Hungarian Forint | - | - | - | - | + | + | + | + | + | + | + | + |
| USDJPY | US Dollar/Japanese Yen | + | + | + | + | + | + | + | + | + | + | + | + |
| USDMXN | US Dollar/Mexican Nuevo Peso | - | - | - | - | + | + | + | + | + | + | + | + |
| USDNOK | US Dollar/Norwegian Krone | + | + | + | + | + | + | + | + | + | + | + | + |
| USDPLN | US Dollar/Polish Zloty | - | - | - | - | + | + | + | + | + | + | + | + |
| USDRUB | US Dollar/Russian Ruble | - | - | - | - | + | + | + | + | + | + | + | + |
| USDSEK | US Dollar/Swedish Krona | + | + | + | + | + | + | + | + | + | + | + | + |
| USDSGD | US Dollar/Singapore Dollar | - | + | + | + | + | + | + | + | + | + | + | + |
| USDTRY | US Dollar/Turkish Lira | - | - | - | - | + | + | + | + | + | + | + | + |
| USDZAR | US Dollar/South Africa Rand | + | + | + | + | + | + | + | + | + | + | + | + |
| XAGUSD | Silver/US Dollar | + | + | + | + | + | + | + | + | + | + | + | + |
| XAUUSD | Gold/US Dollar | + | + | + | + | + | + | + | + | + | + | + | + |
| ZARJPY | South Africa Rand/Japanese Yen | - | - | - | - | + | + | + | + | + | + | + | + |

Table 1 The investigated data are obtained from the indicated instruments in the indicated years.

| τ (min) | 1 | 5 | 10 | 60 | 100 | 250 |
|---|---|---|---|---|---|---|
| EURUSD y | $4.28 \times 10^{-5}$ | $3.01 \times 10^{-5}$ | $2.28 \times 10^{-5}$ | $2.44 \times 10^{-5}$ | $2.55 \times 10^{-5}$ | $2.53 \times 10^{-5}$ |
| EURUSD 1/β | 304 | 308 | 308 | 304 | 303 | 302 |
| XAGUSD y | $-1.46 \times 10^{-3}$ | $7.42 \times 10^{-4}$ | $4.88 \times 10^{-4}$ | $2.61 \times 10^{-4}$ | $2.39 \times 10^{-4}$ | $2.10 \times 10^{-4}$ |
| XAGUSD 1/β | 12.5 | 26.4 | 35.4 | 63.3 | 69.8 | 79.2 |

Table 2 The mean (y) and standard deviation (1/β) values for the empirical distributions of the available data (see, Table 1) from the asserted forex instruments.

| parity | EURUSD | | EURJPY | | GBPUSD | | USDJPY | |
|---|---|---|---|---|---|---|---|---|
| ars | 0.99904 | | 0.99968 | | 0.99921 | | 0.99934 | |
| property | C | α | C | α | C | α | C | α |
| value | 8.99124 | 1.4481 | 8.93249 | 1.4624 | 9.02088 | 1.44562 | 8.95906 | 1.44871 |
| se | 0.06407 | 0.02003 | 0.03733 | 0.01167 | 0.058 | 0.01813 | 0.05343 | 0.01671 |
| | | | | | | | | |
| parity | EURTRY | | USDTRY | | XAGUSD | | XAUUSD | |
| ars | 0.99914 | | 0.99782 | | 0.99994 | | 0.99904 | |
| property | C | α | C | α | C | α | C | α |
| value | 8.41786 | 1.40538 | 8.46095 | 1.40524 | 7.97702 | 1.40292 | 8.63618 | 1.44639 |
| se | 0.05882 | 0.01839 | 0.09387 | 0.02935 | 0.01578 | 0.00494 | 0.06407 | 0.02003 |

Table 3 A least squares fitting results for the parameters of Eq. (30); (ars) stands for the adjusted coefficient of determination (Adj. R-Square) and (se) is the standard error.

FIGURES

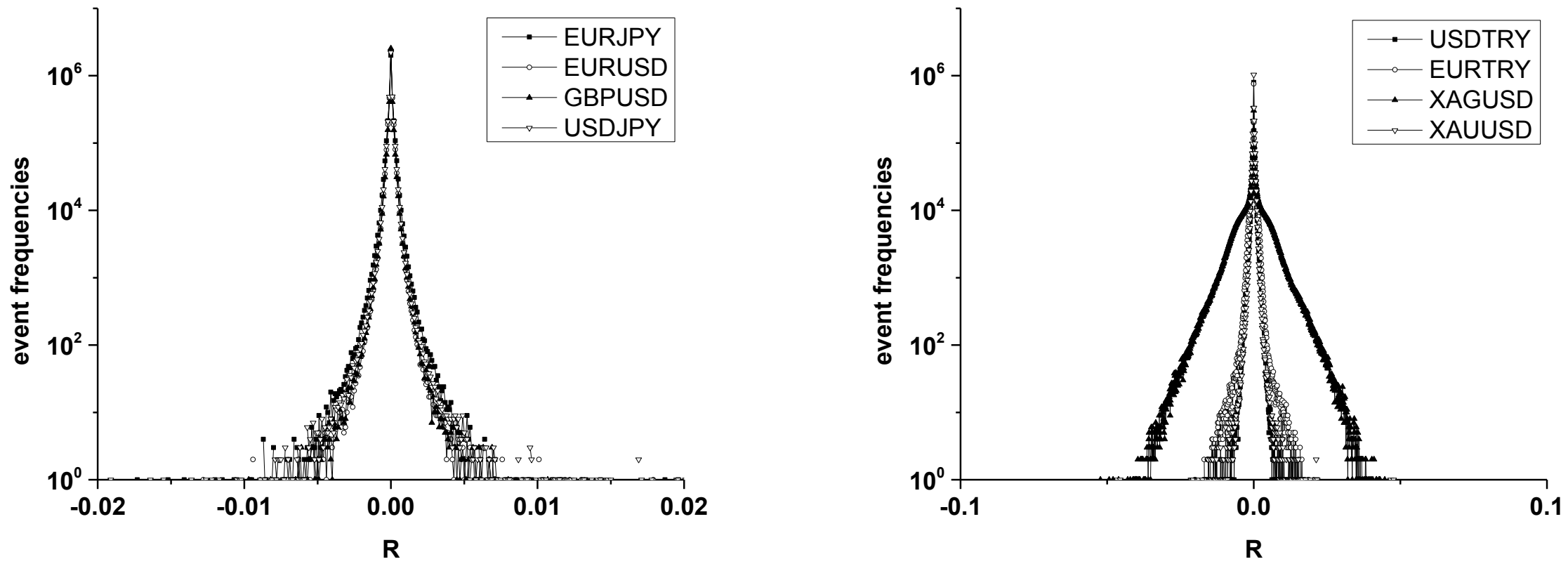

(a) (b)

Figure 1 Empirical distributions of relative returns R(τ=1min;t) obtained from several financial instruments with bin size of 0.0001. Note that the vertical axes are logarithmic.

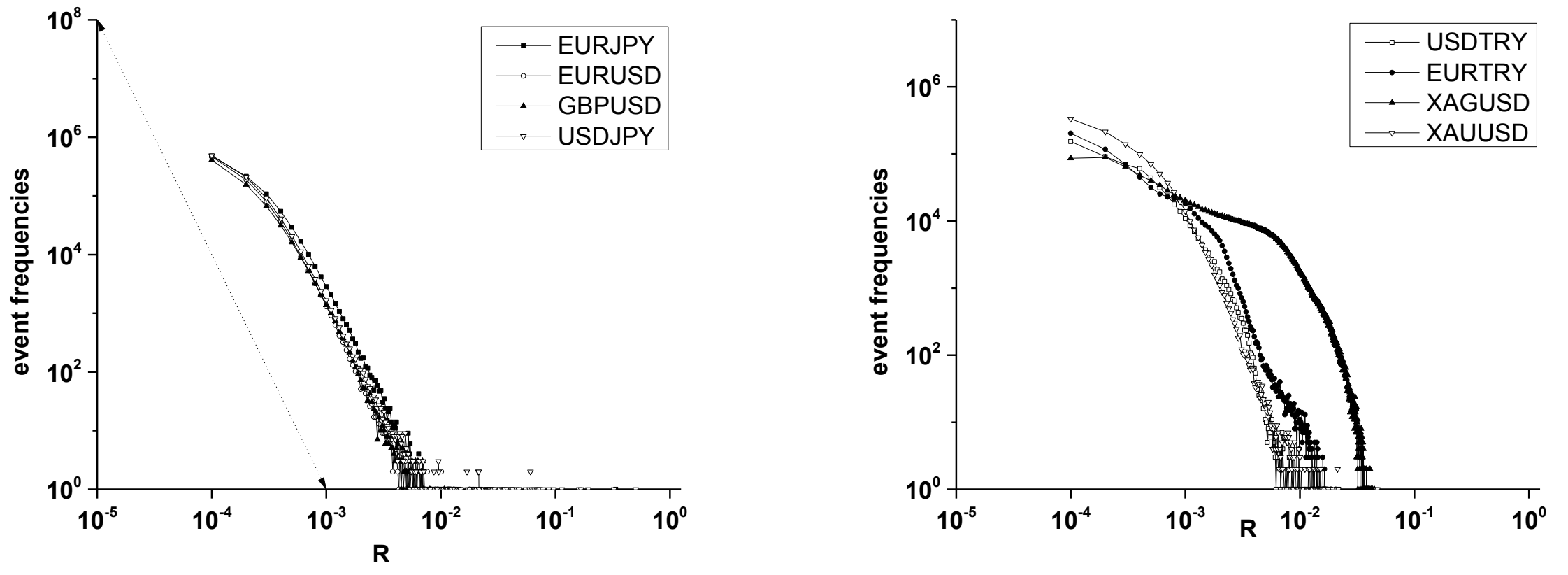

(a) (b)

Figure 2 The same as in Figs. 1 (a) and (b) but the results are presented in log-log scales. Note that the two sided arrow in (a) has a slope around -4 which indicates the index of the power law involved in the related distributions for large magnitudes.

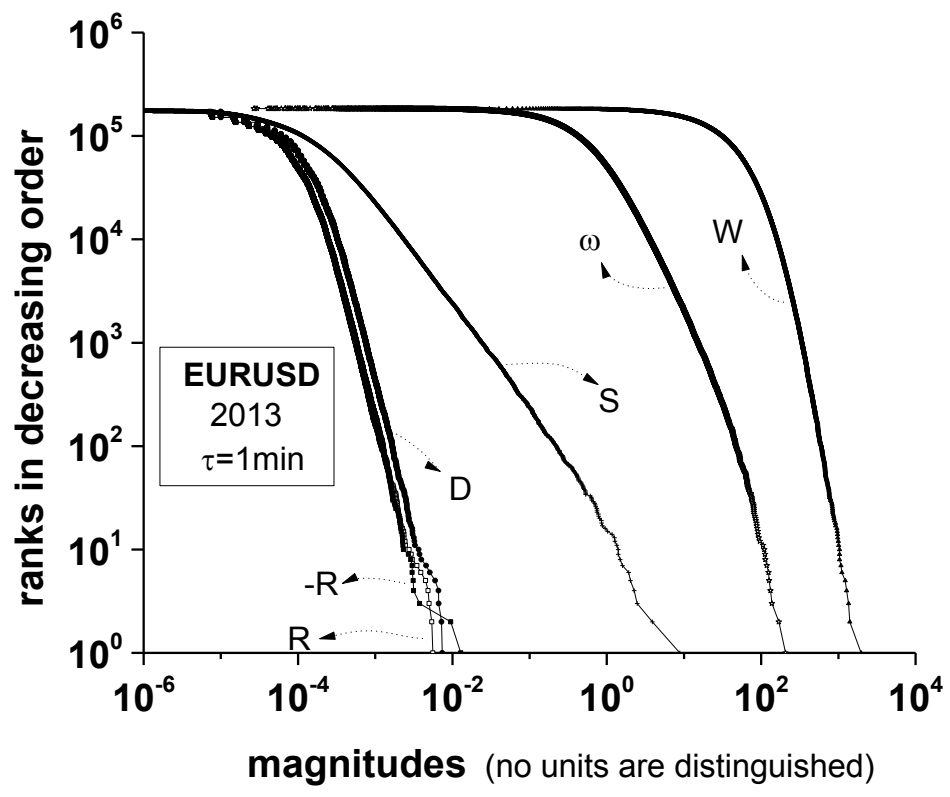

Figure 3 Event frequencies in decreasing rank order plots of the D, W, ω, R, -R and S from the EURUSD parities obtained in the year 2013 with the sampling period of 1 minute. Each line involves 372,312 data points.

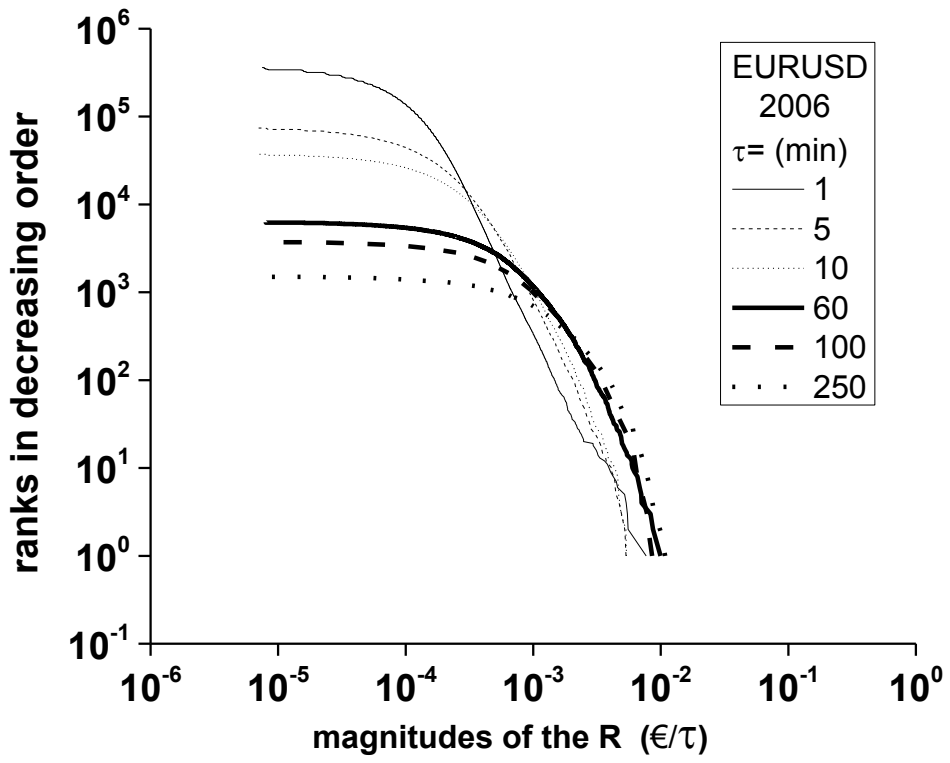

Figure 4 Event frequencies in decreasing rank order plots of the absolute values of the R from the EURUSD parities obtained in the year 2006 with different sampling periods. The line with τ=1 minute involves 374,496 data points.

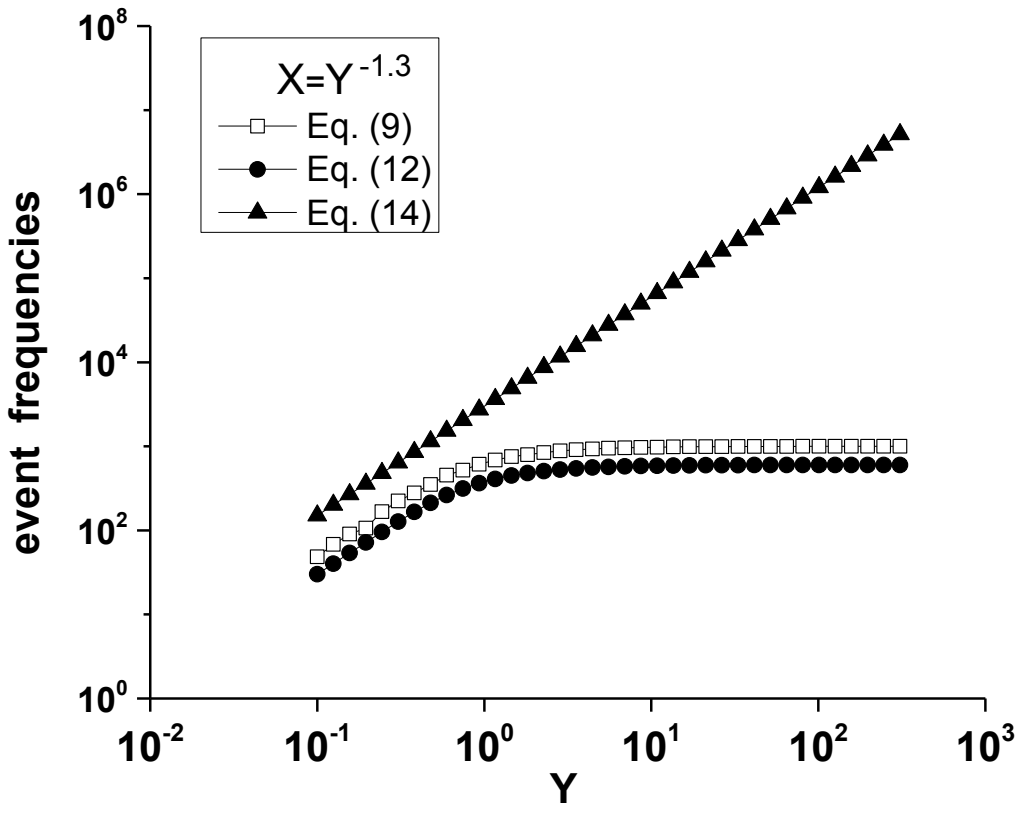

Figure 5 Several simulations for the distribution of sizes $X=Y^{-1.3}$ with J=1000, and $B_{min}=0$ and $B_{max}=1$ in Eq. (9) and (12), respectively.

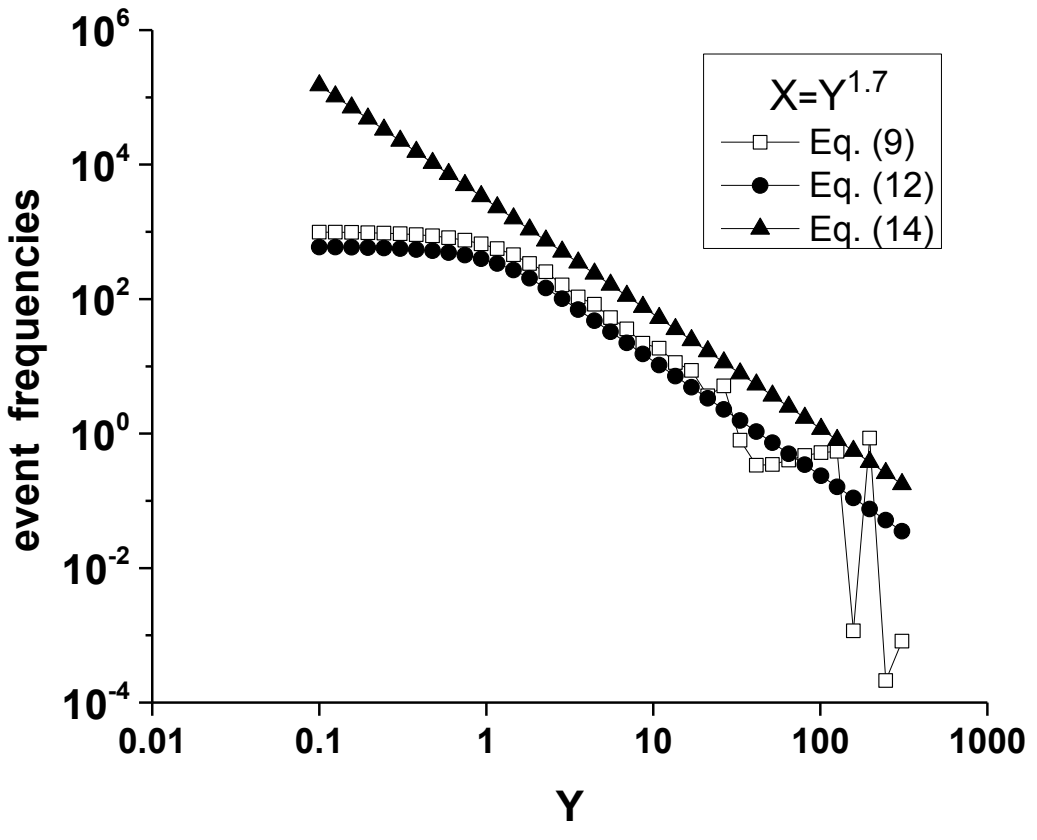

Figure 6 The same as Fig. 5 but for $X=Y^{1.7}$.

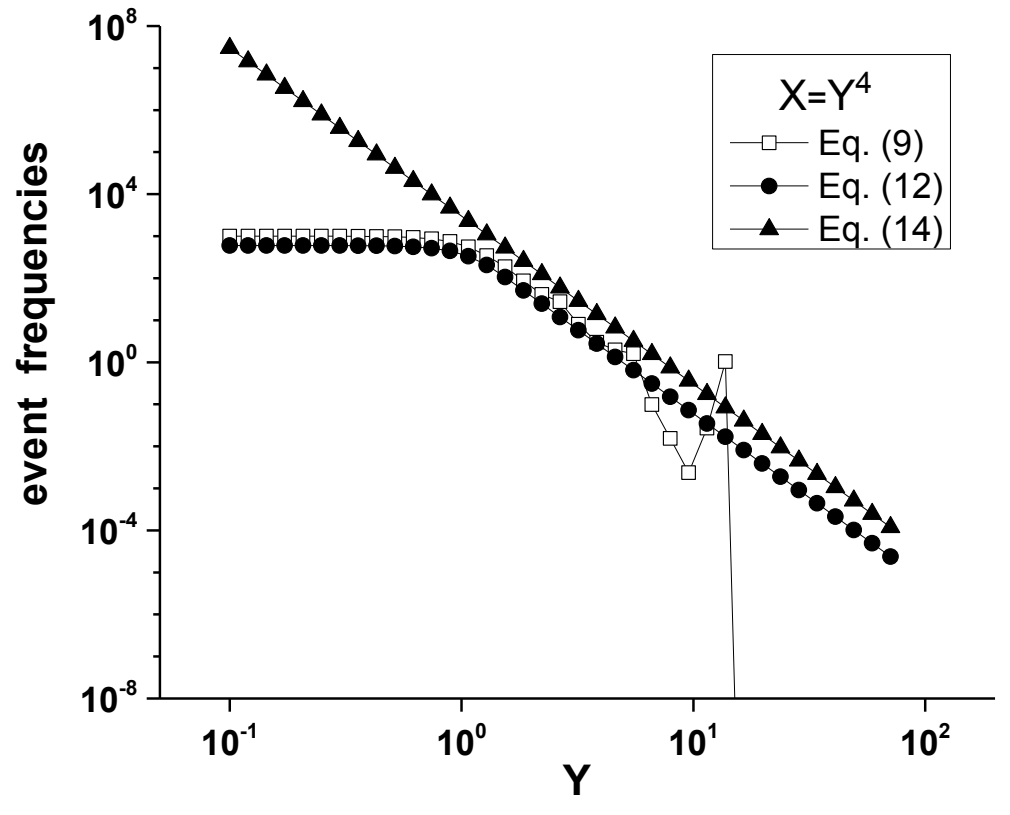

Figure 7 The same as Fig. (5) but for $X=Y^4$.

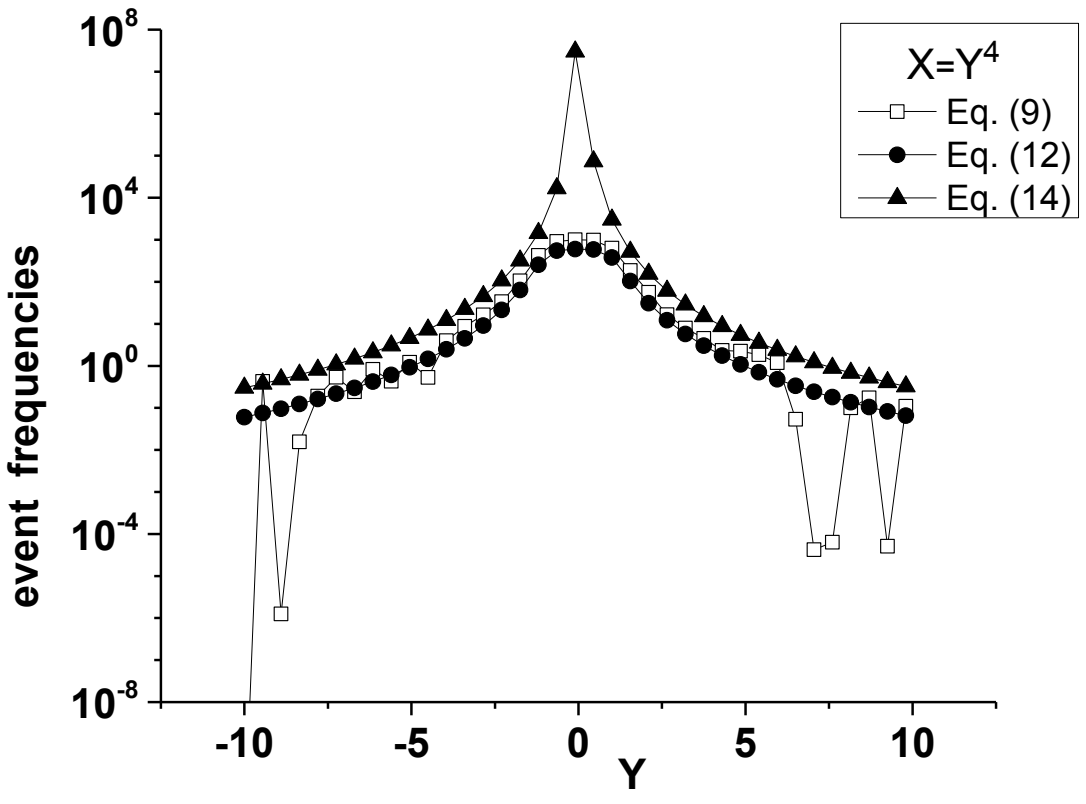

Figure 8 The same as Fig. (7) but in logarithmic scale.

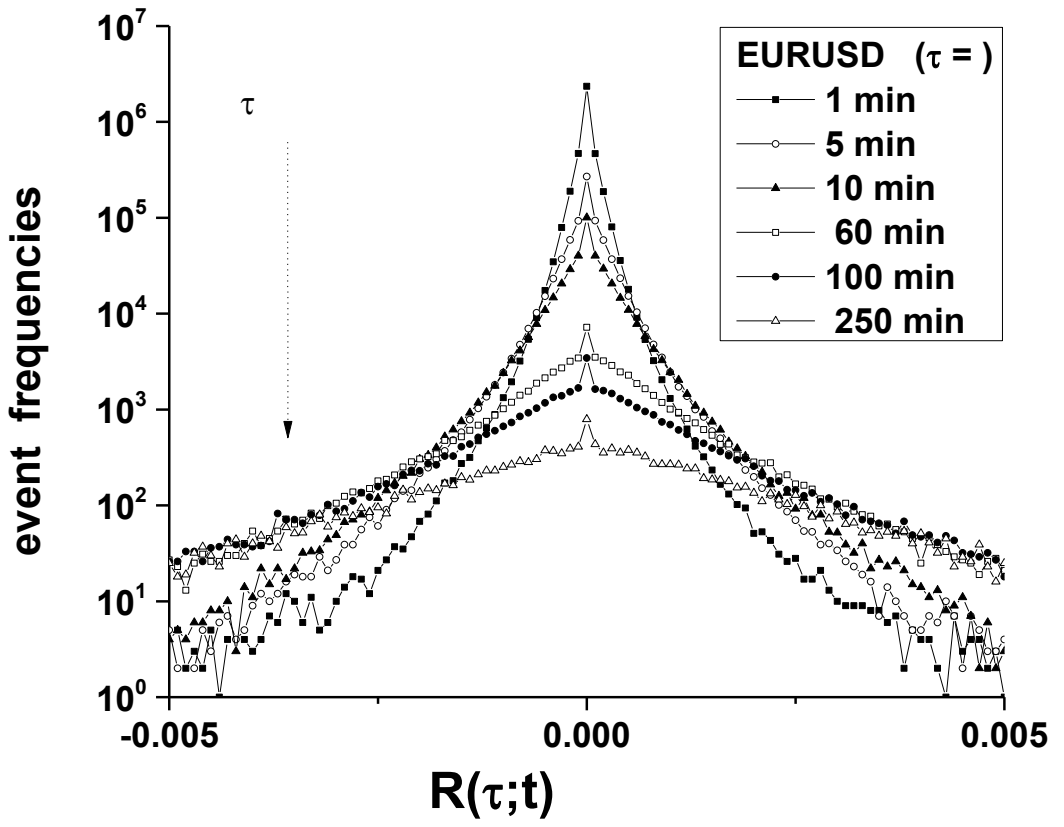

(a)

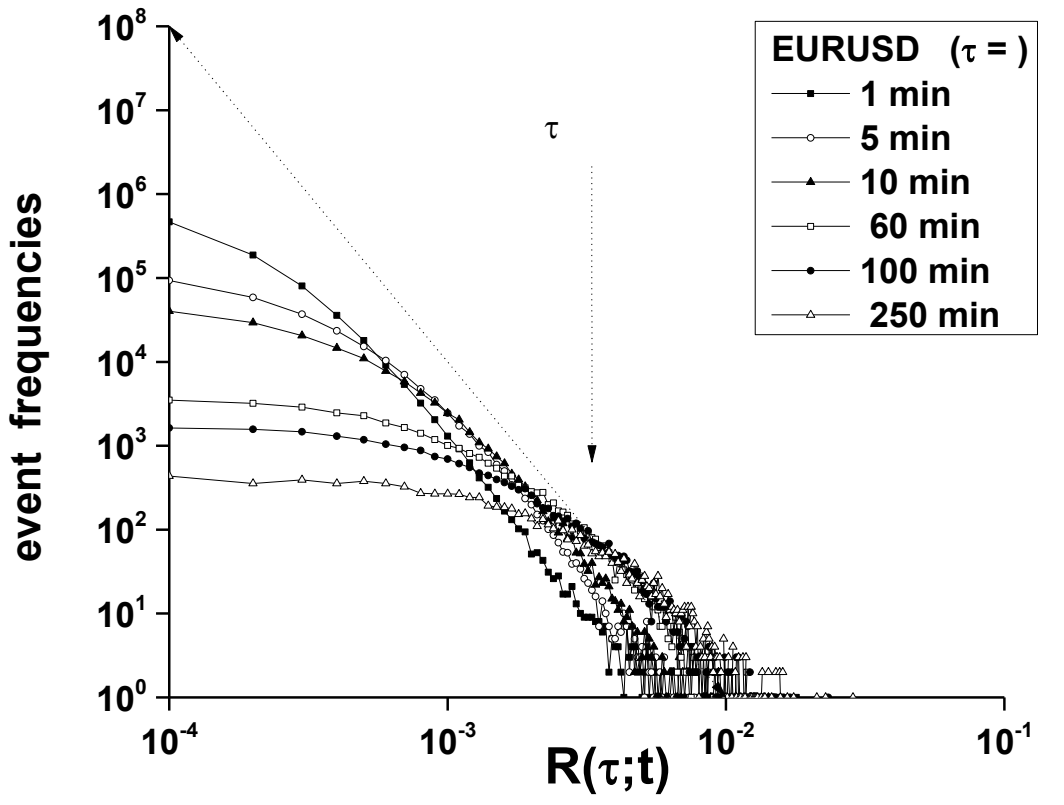

(b)

Figure 9 Empirical distributions of the R(τ;t) from the available data of EURUSD with various sampling periods are exhibited in logarithmic scale in (a) and log-log scale in (b). The used bin sizes commonly equal 0.0001 and the dotted arrows show the direction for the growth of (τ).

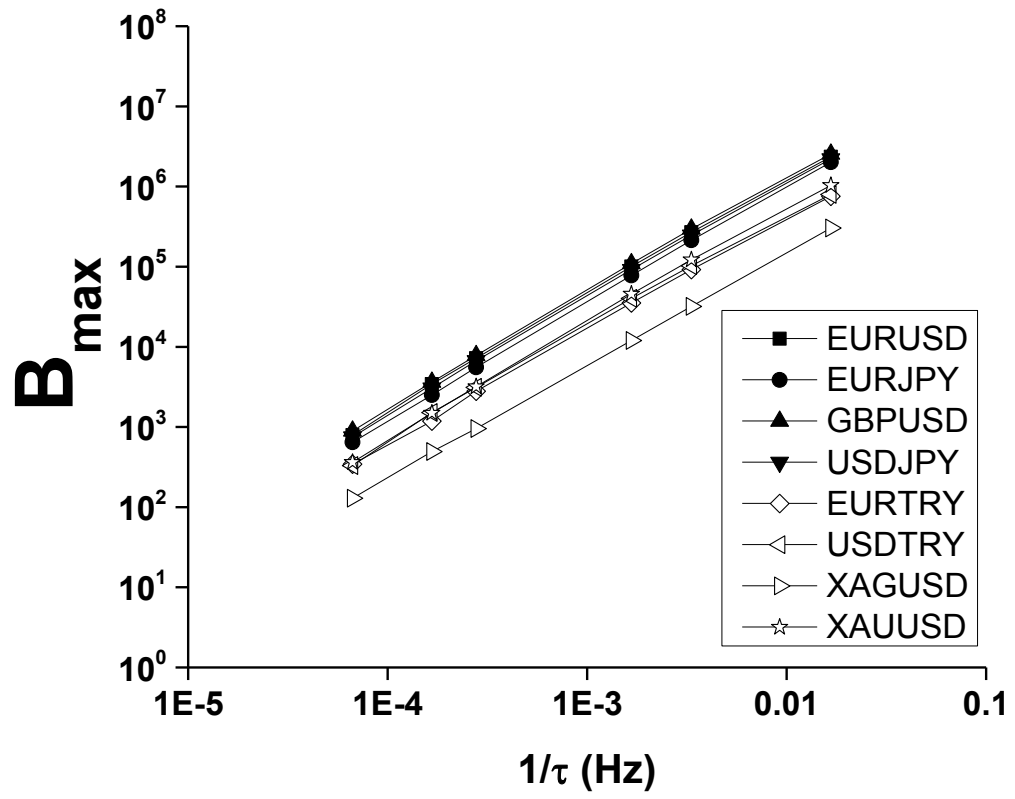

Figure 10 The empirical definition of the values attained by ($B_{max}$) in the empirical distributions of R(τ;t) given in Fig. 1 (a) and (b) as a function of (1/τ) from the asserted forex instruments.

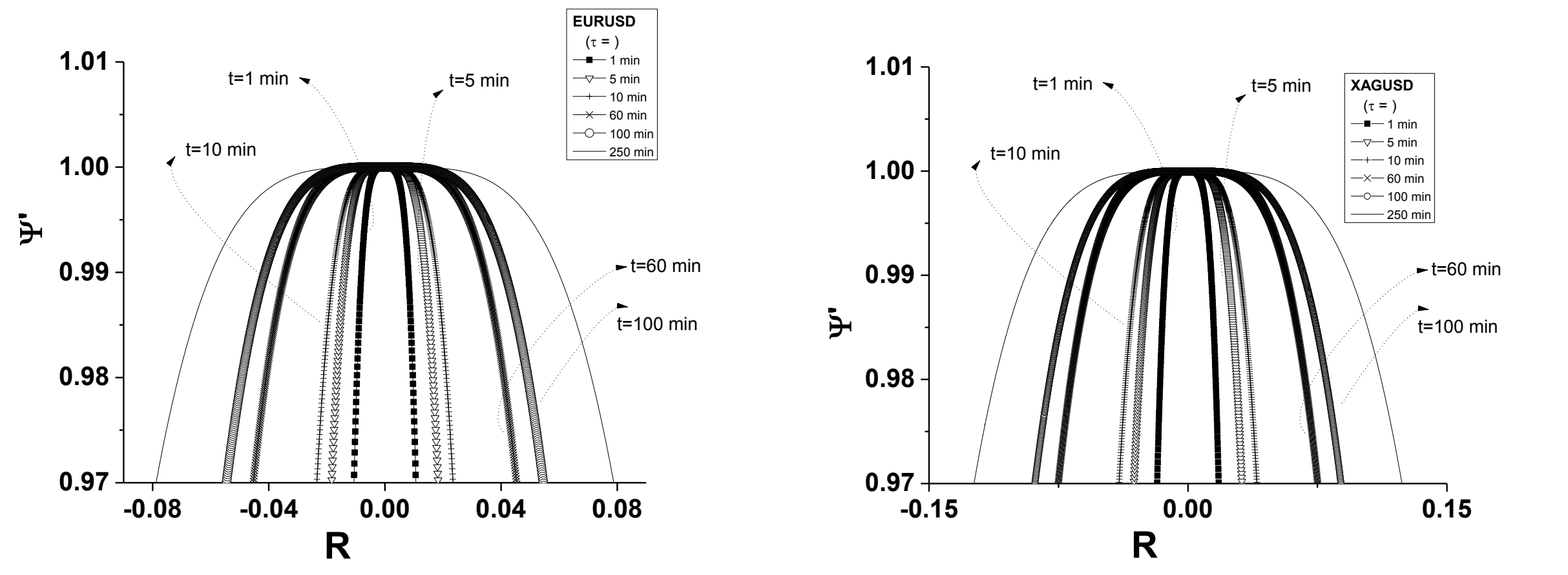

(a) (b)

Figure 11 Data collapsing after Eq.(32) and using the parameters given in Tables 2 and 3 are exemplified for EURUSD and XAGUSD in (a) and (b), respectively.

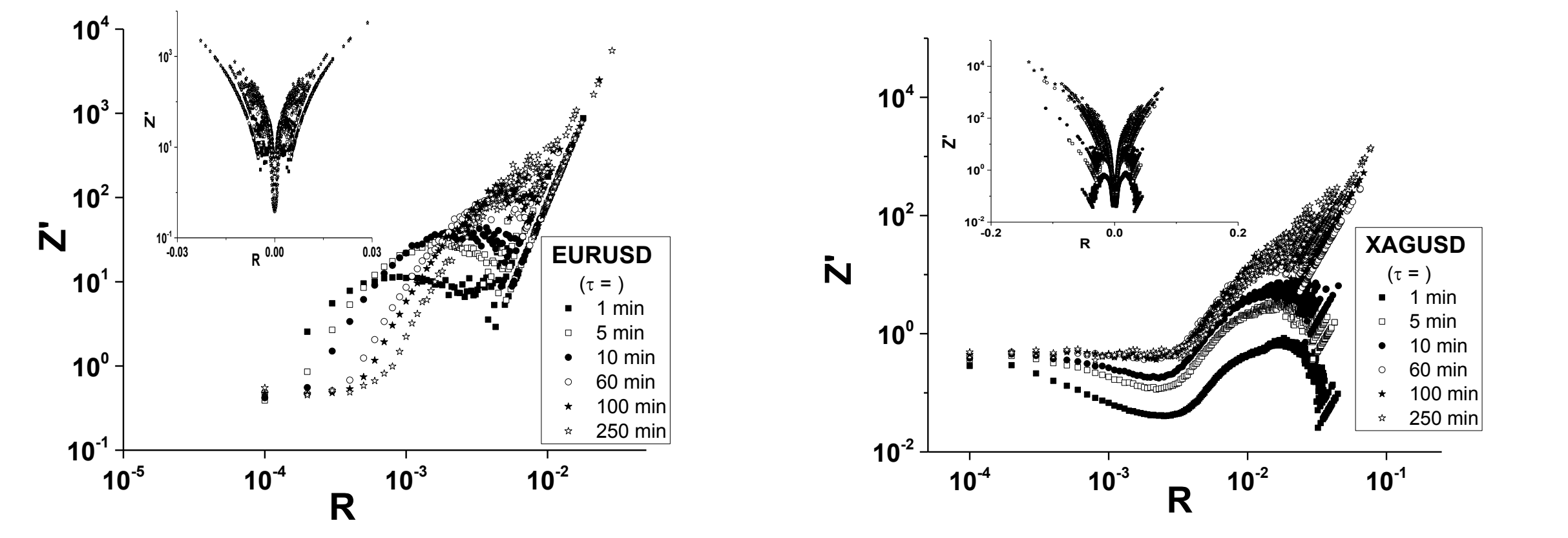

(a) (b)

Figure 12 Data collapsing after Eq.(33) and using the parameters given in Tables 2 and 3 are exemplified in log-log scales for EURUSD and XAGUSD in (a) and (b), respectively. The same plots are displayed in logarithmic scales in the related insets.

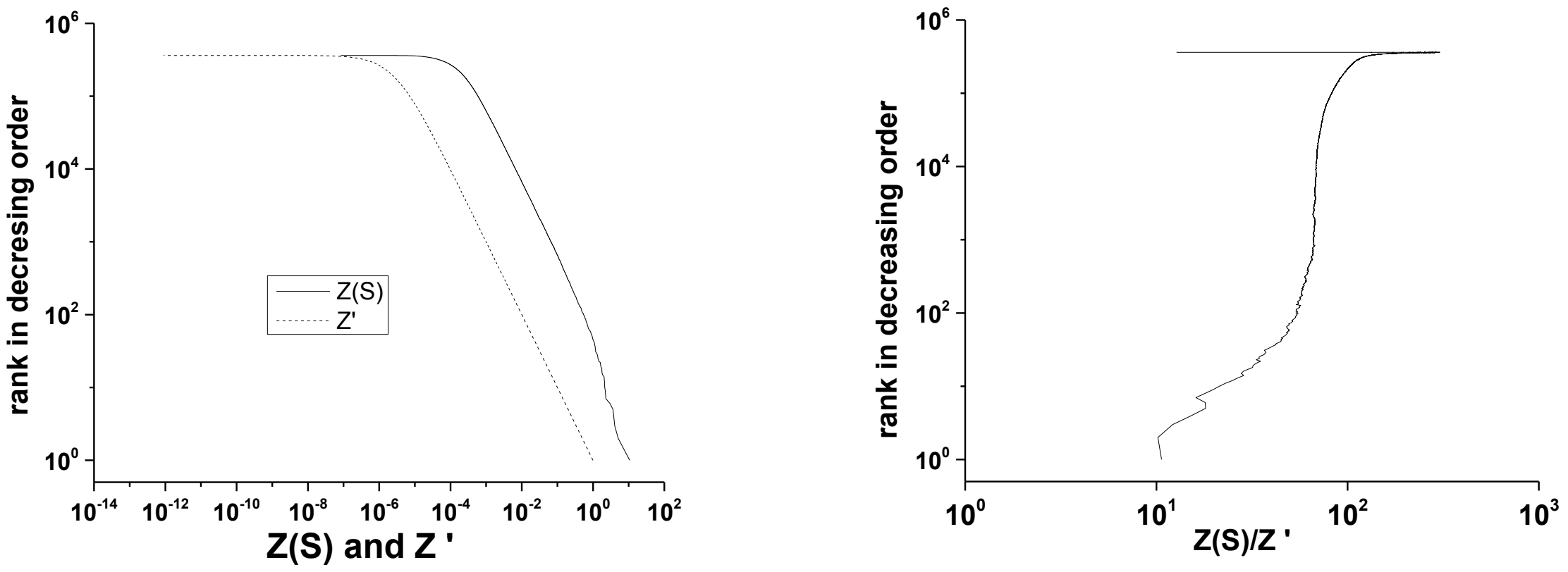

(a) (b)

Figure 13 Approximating the rank distributions of the S(1min,t) from EURUSD in 2006 in terms of that Z' defined in Eq. (29) is displayed in (a) and a measure for the success of that trial is given in (b). The used parameters are y=0, B=1 and 1/β=524,000.

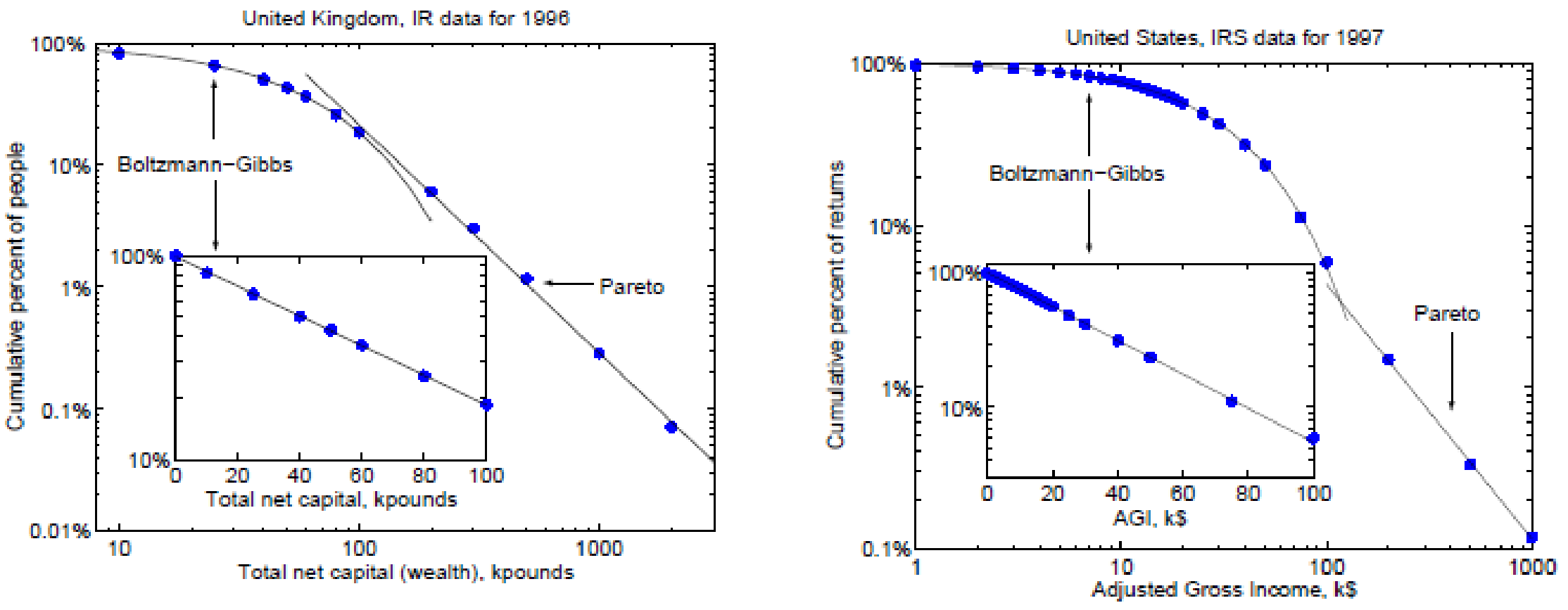

(a) (b)

Figure 14 Cumulative probability distributions of net wealth in the UK are shown in (a) and those of tax returns for USA in 1997 are shown in (b) where the plots are in log-log scales in the main panels and in log-scale in the insets. (From ref [6], used with permission.)

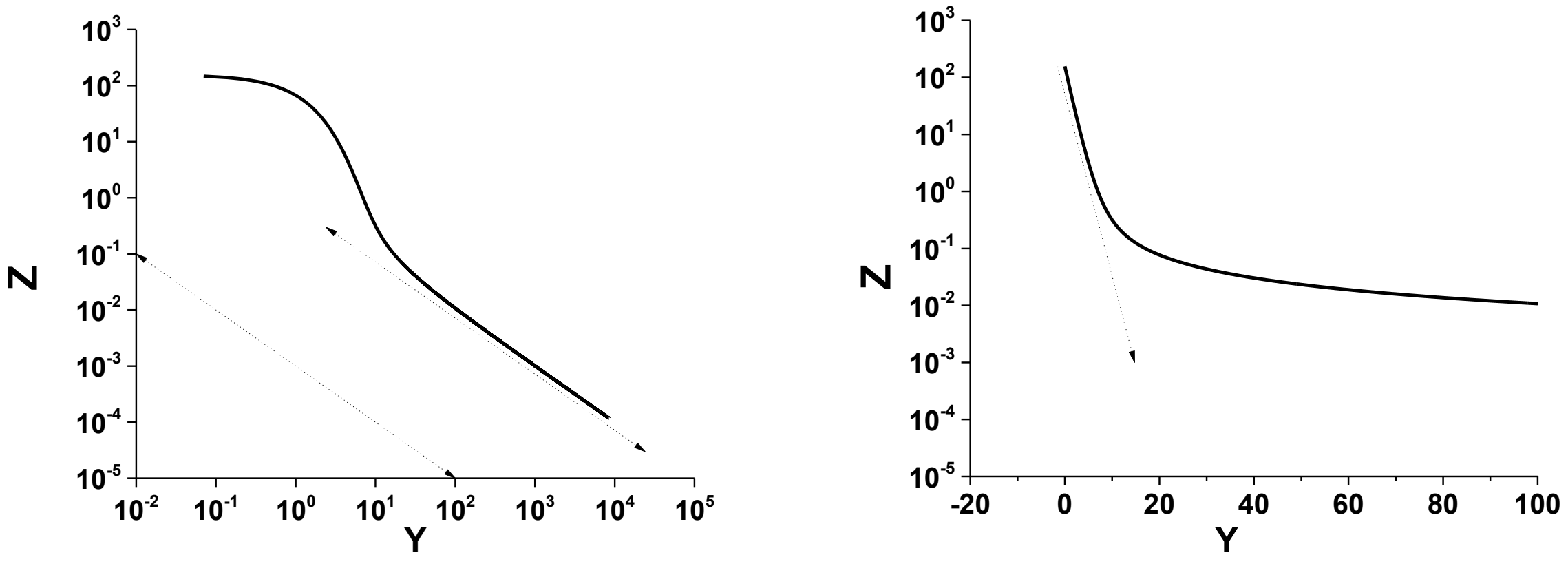

(a) (b)

Figure 15 A simulation for Adjusted Gross Income of the US is depicted in log-log scale (a) and logarithmic scale in (b) where Eq. (28) is utilized with the following parameters; n=1, y=7, $B_{min}$=0, $B_{max}$=1 and β=1. Note that the arrows are parallel in (a) and exponential behavior is clear in (b).